\documentstyle[12pt]{article}

\def\be{\begin{equation}}
\def\ee{\end{equation}}
\def\ben{\begin{displaymath}}
\def\een{\end{displaymath}}
\def\ba{\begin{array}{c}}
\def\bal{\begin{array}{l}}
\def\ea{\end{array}}

\begin{document}

\titlepage
\vspace*{2cm}

 \begin{center}{\Large \bf  Exactly solvable models 
  with  ${\cal PT}-$symmetry 
  
  and with an 
  asymmetric coupling of channels   }\end{center}

\vspace{10mm}

 \begin{center}
Miloslav Znojil

 \vspace{3mm}

\'{U}stav jadern\'e fyziky AV \v{C}R, 250 68 \v{R}e\v{z}, Czech
Republic\footnote{e-mail: znojil@ujf.cas.cz}

\end{center}

\vspace{5mm}

%\today, mulkan.tex

\section*{Abstract}

Bound states generated by the $K$ coupled ${\cal PT}-$symmetric square wells are
studied in a series of models where the Hamiltonians are assumed
${\cal R}-$pseudo-Hermitian and ${\cal R}^2-$symmetric. Specific rotation-like generalized
parities ${\cal R}$ are considered such that ${\cal R}^{N}=I$ at some integers $N$. 
We show that and how our assumptions make
the models
exactly solvable and quasi-Hermitian. This means that they possess the
real spectra as well as the standard probabilistic interpretation.

\vspace{5mm}

PACS

03.65.Ge;
% Solutions of wave equations: bound states
03.65.Ca
% Formalism

\newpage

\section{Introduction}

Bender's and Boettcher's ${\cal PT}-$symmetric version of Quantum
Mechanics~\cite{BB} admits a transition to complex potentials
$V(x)$ (say, on a finite interval) characterized by the ${\cal
PT}-$symmetry property ${\cal PT} V(x) = V(x) {\cal PT}$ where
${\cal P}$ denotes parity while the complex conjugation ${\cal T}$
mimics time reversal. One of the simplest illustrative examples of
the corresponding non-Hermitian ${\cal PT}-$symmetric oscillator
with real spectrum is generated by the purely imaginary
square-well potential step~\cite{sqw}
 \be
 V(x) = V_{(Z)}(x)= - {\rm i}\,Z\,{\rm sign} (x), \ \ \ \ \ \ \
 x \in (-1,1).
 \label{jedenchan}
 \ee
The solvability of this model facilitates an introduction of the
norm of its wave functions in a suitable metric, i.e., via an
introduction of a Hamiltonian-dependent scalar product
\cite{Geyer}. It also renders possible the correct transition to
classical limit~\cite{Batal}. Interesting physical applications of
eq.~(\ref{jedenchan}) were found in supersymmetric
context~\cite{Bagchi} as well as beyond quantum theory where
schematic eq.~(\ref{jedenchan}) and its modifications may play
role in an explanation of the mode-swapping phenomena in classical
magnetohydrodynamics~\cite{Uwe}. In mathematical context, model
(\ref{jedenchan}) helped to clarify the mechanisms of the
spontaneous breakdown of ${\cal PT}-$symmetry at a critical
strength $Z=Z_{crit}$ of non-Hermiticity~\cite{sgezou,Langer}.

A specific merit of model (\ref{jedenchan}) lies in the
feasibility of a transition to its more sophisticated piece-wise
constant solvable alternatives~\cite{two}. A particularly
promising new direction of development has recently been found in
the tentative use of the elementary functions (\ref{jedenchan}) as
forces which mediate an interaction between two~\cite{cc} and/or
three~\cite{nehermpe} coupled square-well oscillators. Here we
intend to move one step further and to re-analyze the similar
coupled $K-$channel problems in a more systematic manner.

Our key idea lies in the observation that in the one-dimensional
Schr\"{o}dinger equation which describes $K$ coupled channels,
 \be
 -\frac{d^2}{dx^2}\varphi^{(m)}(x)
 +
 \sum_{j=1}^{K}\,V_{Z_{(m,j)}}(x)\, \varphi^{(j)}(x)
  =
 E\varphi^{(m)}(x),\ \ \ \ \ m = 1, 2, \ldots, K,
 \label{SEcch}
 \ee
the complications connected with its solution grow very quickly
with $K$. In general, the properties of the model are controlled
by as many as $K^2$ independent real couplings $Z_{(m,j)}$ in
eq.~(\ref{jedenchan}). An introduction of some additional symmetries
would be desirable. Recently we successfully reduced the number of
free parameters to three in the $K=2$ model of ref.~\cite{cc} and,
under the ``stronger" symmetry assumptions, to two in the $K=3$
model of ref.~\cite{nehermpe}.

Inspired by the latter two examples we shall now contemplate $K>3$
channels and try to impose certain symmetry constraints in the
manner which could keep our Schr\"{o}dinger equation (\ref{SEcch})
with more coupled channels exactly and compactly solvable.

%\section{Formulation of the problem}

\section{${\bf PT}-$symmetry revisited}

\subsection{Parity re-interpreted as a pseudo-metric}

In the majority of its updated formulations \cite{psunit,ali,BBJ},
${\cal PT}-$symmetric Quantum Mechanics (PTSQM) replaces the
involutive parity ${\cal P}={\cal P}^{-1}$ by an invertible and
indefinite pseudo-metric ${\bf P}$. In the language coined by Ali
Mostafazadeh \cite{ali} one replaces the ${\cal PT}-$symmetry
property of the Hamiltonian $H \neq H^\dagger$ by the requirement
 \be
 H^\dagger = {\bf P}\,H\,{\bf P}^{-1}\,, \ \ \ \ \
  \ \ {\bf P} = {\bf P}^\dagger \neq I.
 \label{pseudo}
 \ee
It may be understood as a certain necessary condition that the
spectra of the observable $H$ and of its ``redundant" conjugate
$H^\dagger$ coincide.

A ``hidden" purpose of the postulate (\ref{pseudo}) lies in the
fact that as long as $H \neq H^\dagger$, the standard knowledge of
the (presumably, real and discrete) energies $E_n$ and of the
related eigenstates $|n\rangle$ of $H$ must be complemented by the
{\em independent} construction of the eigenstates of the conjugate
operator $H^\dagger$, i.e., in our adapted Dirac's notation, of
the ``ketkets" $|n\rangle\rangle\ $. Fortunately, in such a
situation eq.~(\ref{pseudo}) enables us to employ, in the
non-degenerate case, the implication
 \be
 H^\dagger\,|n\rangle\rangle= E_n\,|n\rangle\rangle
 \ \ \ \ \ \
 \Longrightarrow
 \ \ \ \ \ \
 |n\rangle\rangle = const(n)\,{\bf P}\,|n\rangle\,.
 \label{sepe}
 \ee
The latter condition is of paramount importance for the technical
feasibility of the practical applications of the formalism. 
Indeed, the construction of $|n\rangle\rangle$ becomes straightforward whenever
the action of the pseudo-metric ${\bf P}$ is not too complicated.

A more
detailed support of the latter argument may be found, e.g., in the
Appendices of ref.~\cite{cc} and in ref.~\cite{KG} where we
studied an application of PTSQM to the Peano-Baker-like
two-channel version of Klein-Gordon equation. On the background of this
technical summary we only have to emphasize that the
knowledge of the two sets of the vectors $|n\rangle$ and
$|n\rangle\rangle\ $ (forming a biorthogonal basis in our Hilbert
space) opens, for all the square-well-type models, the way towards
the construction of the necessary ``physical" positive definite
metric $\Theta=\Theta^\dagger$. A nice explicit illustration of
the recipe (which ascribes an appropriate
probabilistic interpretation to the system, cf. Appendix A below) 
has been discussed by Mostafazadeh
and Batal~\cite{Batal}, with an elegant Krein-space mathematical
re-interpretation added recently by Langer and
Tretter~\cite{Langer}.

\subsection{Unitary alternatives to the parity}

We feel inspired by the observation that in a perceivable contrast
to the physical metric $\Theta$ itself, the pseudo-metric plays
just an auxiliary role, via eq. (\ref{sepe}). In such a context,
the requirement of the Hermiticity of ${\bf P}$ is redundant and
the main emphasis must be put on its simplicity. This is the key
idea of our present paper. In place of the standard Hermitian
pseudo-metrics ${\bf P}$ we shall try to work with some
non-Hermitian pseudo-parity operators ${\bf R} \neq {\bf
R}^\dagger$ replacing eq.~(\ref{pseudo}) by its alternative 
 \be
 H^\dagger = {\bf R}\,H\,{\bf R}^{-1}\,, \ \ \ \ \
  \ \ {\bf R} \neq {\bf R}^\dagger.
 \label{psym}
 \label{present}
 \ee
As long as the Hermitian conjugation is an involution, we
may insert eq.~(\ref{present}) in its conjugate version $H = [{\bf
R}^{-1}]^\dagger \,H^\dagger\, {\bf R}^\dagger$ and arrive at the
symmetry requirement
 \be
 H\,{\cal S} = {\cal S}\,H, \ \ \ \ \
 {\cal S} = [{\bf R}^{-1}]^\dagger  {\bf R}.
 \label{symmetry}
 \ee
We believe that its implementation might serve our present
purposes. 

For the sake of definiteness we shall pay attention to
the families of operators ${\bf R}$,
 \ben
 {\bf R}_{(K,1)} =\left (
 \begin{array}{ccccc}
 0&\ldots &0&0&{\cal {P}}\\
 {\cal {P}}&0&\ldots&0&0\\
 0&{\cal {P}}&0&\ldots&0 \\
 \vdots &\ddots&\ddots&\ddots&\vdots \\
 0&\ldots &0&{\cal {P}}&0
 \ea
 \right ), \ \ \ \
 {\bf R}_{(K,2)} =\left (
 \begin{array}{ccccc}
 0&\ldots &0&{\cal {P}}&0\\
 0&0&\ldots &0&{\cal {P}}\\
 {\cal {P}}&0&0&\ldots&0\\
 \vdots &\ddots&\ddots&\ddots& \vdots\\
 0&\ldots &{\cal {P}}&0&0
 \ea
 \right ), \ldots
 \een
with the property ${\bf R}^{-1} = {\bf R}^\dagger$ giving the
symmetry $ {\cal S} = {\bf R}^2$. In the representation where we separate the parity,
${\bf R}_{(K,L)}={\cal {P}}\,{\bf r}_{(K,L)}$, we may employ the
recurrences
 \ben
 {\bf r}_{(K,L+1)} ={\bf r}_{(K,1)}\, {\bf
 r}_{(K,L)}\,,
 \ \ \ \ \ \
 \ \ \ \ \ \
 \ \ \ \ \ \ L =1, 2, \ldots\,.
 \een
At any $K$ the pseudoparities obey the rule $\left [ {\bf r}_{(K,L)}\right
]^K =I$ and may be interpreted as finite rotations. At the even $K =
2M$ we get the non-Hermitian ${\bf r}_{(K,K-L)}=\left [ {\bf
r}_{(K,L)}\right ]^\dagger$ at $L=1, 2, \ldots, M-1$. An anomaly
occurs at $L=M$ where we note that ${\bf r}_{(K,K-L)}=\left [ {\bf
r}_{(K,L)}\right ]^\dagger={\bf r}_{(2M,M)}$ remains Hermitian.
There is no similar Hermitian exception at the odd integers $K =
2M-1$ with $M> 1$.

We are now prepared to apply the rule (\ref{psym}) to our coupled-channel models~(\ref{SEcch}). 

\section{The method}

\subsection{The symmetry-compatible sets of coupling constants}

In units $\hbar = 2m = 1$ and in a partitioned matrix notation
equation (\ref{SEcch}) may be reformulated as a diagonalization of
the Hamiltonians $H = H_{(kinetic)} + H_{(interaction)}(x)$ where
 \ben
 H_{(kinetic)} =
 \left (
 \begin{array}{cccc}
 -\frac{d^2}{dx^2}&0&0&0\\
0&
 -\frac{d^2}{dx^2}&0&0\\
 0&0&\ddots&0\\
 0&0& 0&
 -\frac{d^2}{dx^2}
 \ea
 \right ).
 \een
The separate parity-antisymmetric imaginary square-well potentials
(\ref{jedenchan}) will form the array
 \be
 H_{(interaction)}(x)=
 \left (
 \begin{array}{cccc}
 V_{Z_{(1,1)}}(x)&V_{Z_{(1,2)}}(x)&\ldots&V_{Z_{(1,K)}}(x)\\
 V_{Z_{(2,1)}}(x)&V_{Z_{(2,2)}}(x)&\ldots&V_{Z_{(2,K)}}(x)\\
 \vdots&\vdots&\ddots&\vdots\\
 V_{Z_{(K,1)}}(x)&V_{Z_{(K,2)}}(x)&\ldots&V_{Z_{(K,K)}}(x)
 \ea
 \right )\,
 \label{vazanekanaly}
 \ee
characterized by $K^2$ different real coupling constants,
 \be
   {\bf A}=
 \left (
 \begin{array}{cccc}
 {Z_{(1,1)}}&{Z_{(1,2)}}&\ldots&{Z_{(1,K)}}\\
 {Z_{(2,1)}}&{Z_{(2,2)}}&\ldots&{Z_{(2,K)}}\\
 \vdots&\vdots&\ddots&\vdots\\
 {Z_{(K,1)}}&{Z_{(K,2)}}&\ldots&{Z_{(K,K)}}
 \ea
 \right ).
 \label{vazanekanalyres}
 \ee
As long as the diagonal $H_{(kinetic)}$ commutes with all our
pseudoparities, we shall only have to study the consequences of
the symmetry (\ref{symmetry}) upon the variability of the
set~(\ref{vazanekanalyres}).

In a more detailed analysis of the latter point we must recollect,
firstly, the action of parity ${\cal {P}}$ on our Hamiltonian,
 \ben
  {\cal {P}}\,H_{(interaction)}\,{\cal {P}} =
  -H_{(interaction)}\,.
 \een
This rule enables us to re-write eq.~(\ref{present}) as a
condition imposed upon the real matrix of
indices~(\ref{vazanekanalyres}),
 \be
 {\bf A}= {\bf r}_{(K,K-L)}\cdot {\bf
A}^T\cdot {\bf r}_{(K,L)}\,
 \label{hermiti}
 \ee
where $^T$ denotes transposition. This equation is a core of our
forthcoming constructions. At any fixed $K$ and $L$, it must be
satisfied as a guarantee that our square-well Hamiltonian $H$
obeys the ${\bf R}-$pseudo-Hermiticity rule (\ref{present}).
Step-by-step we shall list the solutions  of eq.~(\ref{hermiti})
distinguishing between the odd $K$ (in a series starting at
section \ref{oddka}) and even $K$ (starting from section
\ref{evenka}).

\subsection{The determination of the bound-state energies
\label{3.22}
}

Our next step may be guided by the elementary single-channel
example of ref.~\cite{sqw} with the ``effective" Schr\"{o}dinger
equation
 \be
 -\frac{d^2}{dx^2}\,\varphi(x)
  - {\rm i}\,Z_{ef\!f}\,{\rm sign} (x)\, \varphi(x)=
 E\,\varphi(x), \ \ \ \ \ \ \ \ \ \
 \varphi(-1) = \varphi(1)=0\,.
 \label{SEes}
 \ee
At $K=1$ and $Z_{ef\!f}=Z$ it has been shown solvable and physical
(i.e., possessing the real spectrum) at $|Z_{ef\!f}| < Z_{crit}
\approx 4.48$ in ref.~\cite{sqw}, with a better estimate of
$Z_{crit} \approx 4.47530856\ $ derived in ref.~\cite{nehermpe}.

For the generic $K > 1$ and for the negative or positive
coordinate $x$, our coupled set (\ref{SEcch}) are differential
equations with constant coefficients. In a way resembling
eq.~(\ref{SEes}) these equations remain solvable by the
trigonometric ansatz
 \be
 \ba
  \varphi^{(m)}(x)
 = \left \{
 \begin{array}{ll}
  C_L^{(m)}\,\sin \kappa_L(x+1)
 , \ \ & x \in (-1,0), \\
  C_R^{(m)}\,\sin \kappa_R(-x+1),
  \ \ & x \in (0,1),\ \ \ \ \ \ \ \ m = 1, 2, \ldots, K
 \ea
 \right .
 \ea
 \label{ansatzf}
 \ee
compatible with the ``external" boundary conditions at $x = \pm
1$. We must also impose the $2K-$plet of the standard ``internal"
matching conditions in the origin,
 \be
 \bal
  C_L^{(m)}\,\sin \kappa_L=
  C_R^{(m)}\,\sin \kappa_R,
 \ \ \ \ \ \ \ \ \ m = 1, 2, \ldots, K,\\
   \kappa_L\,C_L^{(n)}\,\cos \kappa_L=
 -  \kappa_R\,C_R^{(n)}\,\cos \kappa_R,
 \ \ \ \ \ \ \ \ \ n = 1, 2, \ldots, K.
 \ea
 \label{rowsy}
 \ee
Their first half may be read as determining, say, the $K$
``dependent" constants $C_R^{(m)}$ as functions of the $K$
``independent" constants $C_L^{(m)}$ and of the not yet specified
parameters $\kappa_{L,R}$. The ratio of the equations with $m=n$
eliminates all the constants and leads to the single complex
condition
 \be
   \kappa_L\,{\rm cotan}\, \kappa_L=
 -  \kappa_R\,{\rm cotan}\, \kappa_R.
 \label{rowsyx}
 \ee
Finally, the insertion of our ansatz (\ref{ansatzf}) in the
Schr\"{o}dinger equation (\ref{SEcch}) gives 
 \be
 \left (
 \begin{array}{cccc}
  \kappa^2_L+{\rm i}\,Z_{(1,1)}&{\rm i}\,Z_{(1,2)}&\ldots&{\rm i}\,Z_{(1,K)}\\
 {\rm i}\,Z_{(2,1)}& \kappa^2_L+{\rm i}\,Z_{(2,2)}&\ldots&{\rm i}\,Z_{(2,K)}\\
 \vdots&\ddots&\ddots&\vdots\\
 {\rm i}\,Z_{(K,1)}&{\rm i}\,Z_{(K,2)}&\ldots& \kappa^2_L+{\rm i}\,Z_{(K,K)}
 \ea
 \right )\,
 \left (
 \ba
 C_L^{(1)}\\
 C_L^{(2)}\\
 \vdots\\
 C_L^{(K)}
 \ea
 \right )
 =
 E\,\left (
 \ba
 C_L^{(1)}\\
 C_L^{(2)}\\
 \vdots\\
 C_L^{(K)}
 \ea
 \right )
 \label{SEma}
 \ee
at $x \in (-1,0)$ and the similar $K-$dimensional diagonalization
 \be
 \left (
 \begin{array}{cccc}
  \kappa^2_{R}-{\rm i}\,Z_{(1,1)}&-{\rm i}\,Z_{(1,2)}&\ldots&-{\rm i}\,Z_{(1,K)}\\
 -{\rm i}\,Z_{(2,1)}& \kappa^2_{R}-{\rm i}\,Z_{(2,2)}&\ldots&-{\rm i}\,Z_{(2,K)}\\
 \vdots&\ddots&\ddots&\vdots\\
 -{\rm i}\,Z_{(K,1)}&-{\rm i}\,Z_{(K,2)}&\ldots& \kappa^2_{R}-{\rm i}\,Z_{(K,K)}
 \ea
 \right )\,
 \left (
 \ba
 C_{R}^{(1)}\\
 C_{R}^{(2)}\\
 \vdots\\
 C_{R}^{(K)}
 \ea
 \right )
 =
 E\,\left (
 \ba
 C_{R}^{(1)}\\
 C_{R}^{(2)}\\
 \vdots\\
 C_{R}^{(K)}
 \ea
 \right )
 \label{SEna}
 \ee
at $x \in (0,1)$. For the real energies the latter two sets are
just complex conjugates of each other so that we may omit one of
them and  set $\kappa_R = s + {\rm i}t = \kappa_L^*$ with $s > 0$
and $t \in (-\infty,\infty)$.

In the next step generalizing the experience gained in
refs.~\cite{sqw,cc,nehermpe} we use another ansatz
 \be
 E = s^2-t^2.
 \label{ansatzfg}
 \ee
Its use leaves the matrices in eqs.~(\ref{SEma}) and (\ref{SEna})
purely imaginary so that it is easy to write down their common
secular equation which is real,
 \be
 \det\,
 \left (
 \begin{array}{cccc}
  Z_{(1,1)}-2st&Z_{(1,2)}&\ldots&Z_{(1,K)}\\
 Z_{(2,1)}& Z_{(2,2)}-2st&\ldots&Z_{(2,K)}\\
 \vdots&\ddots&\ddots&\vdots\\
 Z_{(K,1)}&Z_{(K,2)}&\ldots&  Z_{(K,K)}-2st
 \ea
 \right )\, =\,0\,.
 \label{secularbe}
 \ee
It is to be complemented by the complex eq.~(\ref{rowsyx}) which,
in the same notation, degenerates to the semi-trigonometric {\em
and $K-$independent} algebraic formula
 \be
 2s\,\sin 2s + 2t\,\sinh 2t = 0.
 \label{seculara}
 \ee
We may summarize that the pair of the real algebraic equations
(\ref{secularbe}) and (\ref{seculara}) may be expected to specify
the real parameters $s=s_n$, $t=t_n$ and the energies $E=E_n$, $n
= 0, 1, \ldots$ at each particular number of channels $K$.

As long as equation (\ref{seculara}) itself is the same for any
$K$, we may treat it simply as a definition of certain
``universal" curve $t=t_{exact}(s)$. It is worth noting that its
shape carries a lot of resemblance to the half-ovals
 \ben
 t_{auxiliary}(s) = \max(0, -s\,\sin 2s)\,.
 \een
(see ref. \cite{sqw} for a more detailed description of the shape
of the exact curve). 

The former equation
(\ref{secularbe}) defines, in principle, the $K-$dependent
$K-$plet of the real eigenvalues $2st\equiv Z_{ef\!f}$. Thus, once we
determine all of them exactly,  $Z_{ef\!f}=Z_{ef\!f}^{(k)}$, $k = 1,
2, \ldots, K$, we may interpret all the related physical roots
$s=s_n$ and $t=t_n$ (which may be degenerate of course) as the
coordinates of the intersections of the above-mentioned universal
half-oval curve $t=t_{exact}(s)$ with any one of the $K$ much more
elementary hyperbolic curves $t_{hyperbolic}^{(k)}(s)=
Z_{ef\!f}^{(k)}/(2s)$.

\section{The first nontrivial model with odd $K=3$ \label{oddka} }

\subsection{Hermitian choices of pseudo-metrics ${\bf P}$ \label{2.32}}

When we pick up $K=3$ we have a nice opportunity to distinguish
between the parity ${\cal P}$, Hermitian pseudometric ${\bf P}$
and its non-Hermitian generalization ${\bf R}$. In the simplest
Hermitian arrangement we may choose the diagonal partitioned
operator
 \ben
 {\bf P}={\bf P}_{(3,0)} =\left (
 \begin{array}{ccc}
 {\cal {P}}&0&0\\
 0&{\cal {P}}&0\\
 0&0&{\cal {P}}
 \ea
 \right )\ =\ {\bf P}^\dagger\ =\
 {\bf P}^{-1}
 %\label{diagprovazanekanaly}
 \een
and notice that its use in eq.~(\ref{psym}) does not impose {any}
constraint upon the nine one-parametric square-well interactions
  \be
 V_{Z_{(j,k)}}(x) =
 \left \{
 \bal
 {\rm i}\,Z_{(j,k)} , \ \ \ \ \ \
  x \in (-1,0),\ \ \ \ \ \ \\
 -{\rm i}\,Z_{(j,k)} , \ \ \ \ \ \ x \in
 (0,1) ,
 \ea
 \right .
 \ \ \ \ \ \ \ \ \ \ j,k = 1, 2, 3.
 \label{SQW}
 \ee
All the nine coupling constants $Z=Z_{(i,j)}$ remain independent. This leaves 
the corresponding solutions very complicated.
There are no symmetries in the problem, it must be solved more or less
purely numerically. This is, from our present constructive point of view,
a not too interesting situation.

The situation is merely marginally improved by the transition to several other, less trivial
Hermitian pseudo-metrics like
 \ben
 {\bf P}={\bf P}_{(3,1)} =\left (
 \begin{array}{ccc}
 0&0&{\cal {P}}\\
 0&{\cal {P}}&0\\
 {\cal {P}}&0&0
 \ea
 \right )\ =\ {\bf P}^\dagger\ =\
 {\bf P}^{-1}
 %\label{dprovazanekanaly}
 \een
or
 \ben
 {\bf P}={\bf P}_{(3,2)} =\left (
 \begin{array}{ccc}
 0&{\cal {P}}&0\\
 {\cal {P}}&0&0\\
 0&0&{\cal {P}}
 \ea
 \right )\ =\ {\bf P}^\dagger\ =\
 {\bf P}^{-1}.
 %\label{dprovazanekanaly}
 \een
It is easy to show that in the Hermitian cases the partitioned
form of the second powers $\left [ {\bf P}_{(3,1)}\right ]^2
=\left [ {\bf P}_{(3,2)}\right ]^2 =I\,{\cal {P}}^2$ becomes
diagonal. 

\subsection{Non-Hermitian pseudo-parities ${\bf R}$ \label{2.3}}

The emerging availability of the first two simplest non-Hermitian
pseudo-parities should be emphasized at $K=3$,
 \ben
 {\bf R}={\bf R}_{(3,1)} =\left (
 \begin{array}{ccc}
 0&0&{\cal {P}}\\
 {\cal {P}}&0&0\\
 0&{\cal {P}}&0
 \ea
 \right )={\bf R}^\dagger_{(3,2)}, \ \ \ \ \
 {\bf R}^{-1}=\left (
 \begin{array}{ccc}
 0&{\cal {P}}&0\\
 0&0&{\cal {P}}\\
 {\cal {P}}&0&0
 \ea
 \right ),
 %\label{fdprovazanekanaly}
 \een
 \ben
 {\bf R}={\bf R}_{(3,2)} =\left (
 \begin{array}{ccc}
 0&{\cal {P}}&0\\
 0&0&{\cal {P}}\\
 {\cal {P}}&0&0
 \ea
 \right )= {\bf R}^\dagger_{(3,1)}, \ \ \ \ \
 {\bf R}^{-1}=\left (
 \begin{array}{ccc}
 0&0&{\cal {P}}\\
 {\cal {P}}&0&0\\
 0&{\cal {P}}&0
 \ea
 \right )= {\bf R}_{(3,1)}.
% \label{gfdprovazanekanaly}
 \een
In contrast to  the Hermitian cases, we only obtain  the higher-power parallel
rules $\left [ {\bf R}_{(3,1)}\right ]^3 =I\,{\cal {P}}^3$ and
$\left [ {\bf R}_{(3,2)}\right ]^3 =I\,{\cal {P}}^3$ since $\left
[ {\bf R}_{(3,1)}\right ]^2 ={\cal {P}}\,{\bf R}_{(3,2)}$ and
$\left [ {\bf R}_{(3,2)}\right ]^2 ={\cal {P}}\,{\bf R}_{(3,1)}$
in both our genuine non-Hermitian samples.

\subsection{The allowed $Z_{(j,k)}$ for the three coupled channels}

At the unique index $L=1$ our ${\bf R}-$pseudo-Hermiticity
condition (\ref{hermiti}) acquires the linear algebraic form
of a set of equations for the couplings $Z_{(j,k)} \equiv a(j,k)$,
 \ben
 \left (
 \begin{array}{ccc}
 0&0&1\\
 1&0&0\\
 0&1&0
 \ea
 \right )
 \left (
 \begin{array}{ccc}
 {a(1,1)}{}&{a(1,2)}{}&{a(1,3)}{}\\
 {a(2,1)}{}&{a(2,2)}{}&{a(2,3)}{}\\
 {a(3,1)}{}&{a(3,2)}{}&{a(3,3)}{}
 \ea
 \right )
 =\ \ \ \ \ \ \ \ \ \ \ \ \ \ \ \ \ \ \
 \een
 \ben
 \ \ \ \ \ \ \ \ \ \ \ \ \ \ \ \ \ \ \ =
 \left (
 \begin{array}{ccc}
 {a(1,1)}{}&{a(2,1)}{}&{a(3,1)}{}\\
 {a(1,2)}{}&{a(2,2)}{}&{a(3,2)}{}\\
 {a(1,3)}{}&{a(2,3)}{}&{a(3,3)}{}
 \ea
 \right )
 \left (
 \begin{array}{ccc}
 0&0&1\\
 1&0&0\\
 0&1&0
 \ea
 \right ).
 \een
Due to the non-Hermiticity of the pertaining matrix ${\bf r}$,
these relations represent a much more powerful constraint which
leaves just the two coupling constants free and independent. The
equations preserve their form under the transposition so that the
structure of our $K=2$ square-well model remains independent of
$L$,
 \be
 {\bf A}_{(interaction)}=
 \left (
 \begin{array}{ccc}
 Z&X&X\\
 X&Z&X\\
 X&X&Z
 \ea
 \right ), \ \ \ \ \ \ L = 1, 2.
 \label{ctyrka}
 \ee
Now we may return to the Hermitian choices of ${\bf P}= {\bf P}_{(3,1)} $ or ${\bf P}=
{\bf P}_{(3,2)} $ of subsection \ref{2.32} which, obviously, introduced much 
less symmetry. Indeed, the solution of the corresponding nine
linear equations generates just the three nontrivial constaints so
that as many as six coupling constants remain independently
variable. The same tendency survives at the higher $K$. We may conjecture 
that the breakdown ${\bf P} \to {\bf R}$ of the Hermiticity is
connected, definitely, with an enhancement of the symmetry and with the
simplicity of the models.

\subsection{Energy levels}

In the final step of the construction of the bound states at $K=3$
we may follow either the general recipe of section \ref{3.22} or 
the recent detailed presentation of the $K=3$ solutions
in ref. \cite{nehermpe}. In essence, we have to connect the parameters $s$ and $t$
with the ``effective charge"',
 \ben
 2st=Z_{ef\!f}, \ \ \ \ Z_{ef\!f}=Z+F
 \een
where,
in the notation of eq.~(\ref{ctyrka}), the three eligible values of the shift 
$F=F_j$ are to be sought 
as eigenvalues $[F_1,F_2,F_3] = [2\,X,-X,-X]$ of the modified matrix
(\ref{ctyrka}),
 \ben
\left (\begin {array}{ccc} 0&X&X\\\noalign{\medskip}X&0&X
\\\noalign{\medskip}X&X&0\end {array}\right )\,.
 \een
The three respective eigenvectors may be found in ref. \cite{nehermpe} -- here 
our MAPLE software produced
their following simpler alternative sample
 \ben
 \left \{1,1,1
\right \}, \left \{-1,0,1\right \}, \left \{-1,1,0\right \}
 \een
which is still to be re-orthogonalized.

In a way compatible with
the results of ref.~\cite{nehermpe}
we may summarize that our $K=3$ coupled bound states
are determined by formulae
(\ref{ansatzf}) and (\ref{ansatzfg}). The
parameters $s$ and $t$ are fixed as intersections of the half-ovals
(\ref{seculara}) with one of the two available
hyperbolic curves,
  \be
  t=t^{(\sigma)} (s) = \frac{1}{2s}\,Z_{ef\!f}(\sigma), \ \ \ \sigma = 1,2,
  \ \ \
  Z_{ef\!f}(1) = Z+2\,{X}, \ \ \ \ \
  Z_{ef\!f}(2) = Z-{X}\,.
 \label{rooty}
 \ee
By construction, the second family of intersections represents the twice-degenerate levels.

\section{The next model with odd $K=5$}

%\subsection{Pseudoparities and allowed couplings for the five coupled channels}

There is no anomaly in the non-Hermiticity of ${\bf r}={\bf
r}_{(5,L)}$ with $L = 1, 2, 3$ and $4$,
 \ben
 {\bf r}_{(5,1)}=
 \left (\begin {array}{ccccc}
 0&0&0&0&1\\1&0&0&0&0
 \\ 0&1&0&0&0\\ 0&0&1&0&0
 \\ 0&0&0&1&0\end {array}\right ), \ \ \ldots, \ \
 {\bf r}_{(5,4)}=
 \left (\begin {array}{ccccc}
 0&1&0&0&0\\0&0&1&0&0
 \\0&0&0&1&0
 \\0&0&0&0&1
 \\1&0&0&0&0
 \\
 \end {array}\right ).
 \een
All the four different pseudo-Hermiticity conditions
(\ref{hermiti}) with $L = 1,2,3,4$ lead to the same
three-parametric coupling-constant matrix
 \be
 {\bf A}_{(interaction)}=
 \left (
 \begin{array}{ccccc}
 Z&X&D&D& X\\
 X&Z&X&D&D \\
 D&X&Z&X&D \\
 D&D&X&Z&X \\
 X &D &D &X &Z
 \ea
 \right ).
 \label{petidim}
 \ee
Besides its exceptional eigenvector $\left \{1,1,1,1,1 \right \}$
pertaining to the obvious eigenvalue $F_0= 2\,D+2\,X$, the reduced $Z=0$ form 
of this matrix is most easily shown to possesses the pair of the twice
degenerate eigenvalues,
 \ben
 F_{\pm }=\frac{1}{2}\,
 \left [-D-X\pm \sqrt {5}\left (-D+X\right ) \right ]
 \een
with the two respective eigenvectors
 \ben
 \left \{ \frac{1}{2}\,{\frac {
D-X \pm \sqrt {5}\left (-D+X\right )}{-X+D}},
  -\frac{1}{2}\,{\frac {
D-X \pm \sqrt {5}\left (-D+X\right )}{-X+D}},1,0,-1 \right \}
 \een
and
 \ben
 \left \{ \frac{1}{2}\,{\frac
{D-X \pm \sqrt {5}\left (-D+X\right )}{-X+D}},-1,
    0,1, -\frac{1}{2}\,
 \frac{D-X \pm
  \sqrt{5} (-D+X ) }{-X+ D}
  \right \}.
 \een
The next steps towards the next odd $K=7, 9, \ldots$ will be discussed 
in our concluding remarks. Now, for a more specific illustration of some 
technical subtleties let
us return to the systems with the small even numbers of coupled channels.

\section{The simplest model with even $K=2$  \label{evenka} }

It would be easy to
relax the involution assumption ${\cal {P}}^2=I$ as formally
redundant and pedagogically partially misleading. Even without such a 
constraint we get just the most elementary Hermitian
option at $K=2$,
 \ben
 {\bf P} =\left (
 \begin{array}{cc}
 {\cal {P}}&0\\
 0&-{\cal {P}}
 \ea
 \right ) ={\bf P}^\dagger , \ \ \ \ \
 {\bf P}^{-1}=\left (
 \begin{array}{cc}
 {\cal {P}}^{-1}&0\\
 0&-{\cal {P}}^{-1}
 \ea
 \right ),
 %\label{aprovazanekanaly}
 \een
plus its fully off-diagonal alternative
 \ben
 {\bf P} =\left (
 \begin{array}{cc}
 0&{\cal {P}}\\
 {\cal {P}}&0
 \ea
 \right )={\bf P}^\dagger, \ \ \ \ \
 {\bf P}^{-1}=\left (
 \begin{array}{cc}
 0&{\cal {P}}^{-1}\\
 {\cal {P}}^{-1}&0
 \ea
 \right ).
 %\label{bprovazanekanaly}
 \een
Only for the latter sample choice of the
matrix ${\bf r}= {\bf r}_{(2,1)}$ the compactified version
(\ref{hermiti}) of the pseudo-Hermiticity condition
(\ref{present}) acquires a nontrivial two-by-two matrix
form,
 \ben
 \left (
 \begin{array}{cc}
 0&1\\
 1&0
 \ea
 \right )
 \left (
 \begin{array}{cc}
 {a(1,1)}{}&{a(1,2)}{}\\
 {a(2,1)}{}&{a(2,2)}{}
 \ea
 \right )
 =
 \left (
 \begin{array}{cc}
 {a(1,1)}{}&{a(2,1)}{})\\
 {a(1,2)}{}&{a(2,2)}{}
 \ea
 \right )
 \left (
 \begin{array}{cc}
 0&1\\
 1&0
 \ea
 \right ).
 \een
With involutive ${\cal P}$ this model as well as the resulting set of the four equations 
has already been studied in ref.~\cite{cc} and may be easily shown to
degenerate to the single constraint
$a(1,1) = a(2,2)$. Thus, the $K=2$ version of our present
square-well model possesses three free real parameters
$X, Y$ and $Z$,
 \be
 {\bf A}_{(interaction)}=
 \left (
 \begin{array}{cc}
 Z&Y\\
 X&Z
 \ea
 \right ), \ \ \ \ \ \ L = 1.
 \label{trojka}
 \ee
As long as the pseudo-parity remains Hermitian, the model does not
fit in the scope of our present paper. Still it exemplifies the general pattern of the construction
since both the eigenvalues $F=F_{\pm}$ of the modified $Z=0$ version of
matrix (\ref{trojka}) are easily found, $ F_{\pm}=\pm \sqrt {XY}$ and 
also the determination of the two respective eigenvectors 
by our MAPLE program,
 \ben
\left \{1,{ {\sqrt {X/Y}}}\right \},\left \{1,-{
{\sqrt {X/Y}}}\right \}
 \een
is easily verified by hand and tests the recipe.

\section{Four coupled channels \label{2.5}}

Just a smaller representative sample is to be added at $K=4$, viz,
the non-Hermitian
 \ben
 {\bf R}={\bf R}_{(4,1)} =\left (
 \begin{array}{cccc}
 0&0&0&{\cal {P}}\\
 {\cal {P}}&0&0&0\\
 0&{\cal {P}}&0&0\\
 0&0&{\cal {P}}&0
 \ea
 \right )\neq {\bf R}^\dagger, \ \ \ \ \
 {\bf R}^{-1}=\left (
 \begin{array}{cccc}
 0&{\cal {P}}^{-1}&0&0\\
 0&0&{\cal {P}}^{-1}&0\\
 0& 0&0&{\cal {P}}^{-1}\\
 {\cal {P}}^{-1}&0&0&0
 \ea
 \right )
 %\label{fdprovazanekanaly}
 \een
and the ``exceptional" Hermitian
 \ben
 {\bf R}={\bf R}_{(4,2)} =\left (
 \begin{array}{cccc}
 0&0&{\cal {P}}&0\\
 0&0&0&{\cal {P}}\\
 {\cal {P}}&0&0&0\\
 0&{\cal {P}}&0&0
 \ea
 \right )= {\bf R}^\dagger, \ \ \ \ \
 {\bf R}^{-1}=\left (
 \begin{array}{cccc}
 0&0&{\cal {P}}^{-1}&0\\
 0& 0&0&{\cal {P}}^{-1}\\
 {\cal {P}}^{-1}&0&0&0\\
 0&{\cal {P}}^{-1}&0&0
 \ea
 \right ).
 %\label{fdprovazanekanaly}
 \een
It should be noticed that while we obtain a diagonal $\left [ {\bf
R}_{(4,2)}\right ]^2 =I\,{\cal {P}}^2$ in the second power of our
Hermitian operator, and analogous non-Hermitian formula requires
the use of the fourth power, $\left [ {\bf R}_{(4,1)}\right ]^4
=I\,{\cal {P}}^4$.

The overall structure of the  matrix of couplings ${\bf A}$
compatible with our requirement of the ``maximal" non-Hermitian
symmetries (\ref{hermiti}) ceases to be unique at $K=4$. {\it A
priori}, this follows from the observation that besides the
matrix-transposition mapping between the pseudo-Hermiticity
constraints at $L=1$ and $L=3$, one also encounters the anomalous
Hermitian problem at $L=2$. In the latter case we have to solve
the 16 linear algebraic equations
 \ben
 \left (
 \begin{array}{cccc}
 0&0&1&0\\
 0&0&0&1\\
 1&0&0&0\\
 0&1&0&0
 \ea
 \right )
 \left (
 \begin{array}{cccc}
 {a(1,1)}{}&{a(1,2)}{}&{a(1,3)}{}&{a(1,4)}{}\\
 {a(2,1)}{}&{a(2,2)}{}&{a(2,3)}{}&{a(2,4)}{}\\
 {a(3,1)}{}&{a(3,2)}{}&{a(3,3)}{}&{a(3,4)}{}\\
 {a(4,1)}{}&{a(4,2)}{}&{a(4,3)}{}&{a(4,4)}{}
 \ea
 \right )
 =\ \ \ \ \ \ \ \ \ \ \ \ \ \ \ \ \ \ \
 \een
 \ben
 \ \ \ \ \ \ \ \ \ \ \ \ \ \ \ \ \ \ \ =
 \left (
 \begin{array}{cccc}
 {a(1,1)}{}&{a(2,1)}{}&{a(3,1)}{}&{a(4,1)}{}\\
 {a(1,2)}{}&{a(2,2)}{}&{a(3,2)}{}&{a(4,2)}{}\\
 {a(1,3)}{}&{a(2,3)}{}&{a(3,3)}{}&{a(4,3)}{}\\
 {a(1,4)}{}&{a(2,4)}{}&{a(3,4)}{}&{a(4,4)}{}
 \ea
 \right )
 \left (
 \begin{array}{cccc}
 0&0&1&0\\
 0&0&0&1\\
 1&0&0&0\\
 0&1&0&0
 \ea
 \right )
 \een
which only impose the six constraints upon the 16 couplings. From
the practical point of view, too many of them remain freely
variable. Similar results are obtained also for the other
Hermitian operators ${\bf R}$.

In contrast, the parallel and transposition-related $L=1$ and
$L=3$ non-Hermitian versions of eqs.~(\ref{present}) give the set
of 16 equations
 \ben
 \left (
 \begin{array}{cccc}
 0&0&0&1\\
 1&0&0&0\\
 0&1&0&0\\
 0&0&1&0
 \ea
 \right )
 \,{\bf A}
 =
 {\bf A}^T\,
 \left (
 \begin{array}{cccc}
 0&0&0&1\\
 1&0&0&0\\
 0&1&0&0\\
 0&0&1&0
 \ea
 \right ), \ \ \ L=1,
 \een
and
 \ben
 \left (
 \begin{array}{cccc}
 0&1&0&0\\
 0&0&1&0\\
 0&0&0&1\\
 1&0&0&0
 \ea
 \right )
 \,{\bf A}
 =
 {\bf A}^T\,
 \left (
 \begin{array}{cccc}
 0&1&0&0\\
 0&0&1&0\\
 0&0&0&1\\
 1&0&0&0
 \ea
 \right ), \ \ \ L=3,
 \een
respectively. Both of them give {\em the same} four-parametric set
of coupling constants
 \be
 {\bf A}_{(interaction)}=
 \left (
 \begin{array}{cccc}
 Z&U&D&U\\
 L&Z&L&D\\
 D&U&Z&U\\
 L &D &L &Z
 \ea
 \right ), \ \ \ \ \ \ L = 1,3,
 \label{ctyrdim}
 \ee
compatible with both our non-Hermitian pseudo-parities and with
the four quadruplets of the independent elements sitting on the
main diagonal ($Z$), side diagonals ($D$) and in an upper square
($U$) and lower square ($L$). After an obvious permutation of the
basis this matrix may be understood as a partitioned structure
 \be
 {\bf A}_{(interaction)}^{(permuted)}=
 \left (
 \begin{array}{cc|cc}
 Z&D&U&U\\
 D&Z&U&U\\
 \hline
 L&L&Z&D\\
 L &L &D &Z
 \ea
 \right ), \ \ \ \ \ \ L = 1,3,
 \label{dimctyr}
 \ee
which represents an partitioned generalization of the two-by-two
model (\ref{trojka}) above. Directly, this solution may be derived
from the pseudo-parity
 \ben
 {\bf r}^{(permuted)}={\cal {P}}^{-1}{\bf R}^{(permuted)}=
 \left (
 \begin {array}{cccc} 0&0&1&0\\0&0&0&1
 \\ 0&1&0&0\\1&0&0&0\end {array}
 \right )
 \een
which is just a permuted version of the above non-Hermitian square
root ${\bf r}_{(4,1)}$ of the unit matrix.

%\subsection{Bound states }

%\subsubsection{$K=4$, first option}

Four eigenvalues $F_j=
[-D,-D,D+2\,\sqrt {UL},D-2\,\sqrt {UL}]$ of the modified matrix (\ref{ctyrdim}) with $Z=0$
form the partially degenerate quadruplet.
The respective eigevectors read
 \ben
\left \{1,0,-1,0\right \}, \left \{0,1,0,-1\right \},
 \left \{1,{\frac {\sqrt {UL}}{U}},1,{\frac {\sqrt
{UL}}{U}}\right \} ,
 \left \{1,-{\frac {\sqrt
{UL}}{U}},1,-{\frac { \sqrt {UL}}{U}}]\right \}.
 \een
The quadruplet of eigenvalues
derived from the alternative
four-by-four matrix (\ref{dimctyr}) with $Z=0$ remains unchanged of course.
Even the respective eigenvectors themselves become merely predictably
influenced by the underlying permutation,
 \ben
\left \{0,0,1,-1\right \}, \left \{-1,1,0,0 \right \},
 \left \{1,1,{\frac {\sqrt {UL}}{U}},{\frac {\sqrt
{UL}}{U}} \right \},
 %\een
 %\ben
 \left \{1,1,-{\frac
{\sqrt {UL}}{U}},-{\frac { \sqrt {UL}}{U}}\right \}.
 \een

\section{Six coupled channels}

While the choice of the Hermitian pseudo-parity ${\bf r}_{(6,3)}$
is not sufficiently restrictive and leaves 21 free parameters in
the related six-by-six coupling-matrix ${\bf A}$, much more
symmetry (with just 7 free parameters) is induced by all the
non-Hermitian ${\bf r}_{(6,L)}$ with $L \neq 3$.

It is worth noting that different patterns are obtained at $L=1$
or $L=5$ (when the resulting real matrix ${\bf A}$ remains
asymmetric) and at $L=2$ or $L=4$ (when the resulting real matrix
${\bf A}$ becomes symmetric). In the former case, a suitable
permutation of the matrix indices leads to a maximally compact
picture,
 \be
 {\bf r}_{(6,1)}^{(permuted)}=
 \left (\begin {array}{cccccc}
 0&0&0&1&0&0
 \\
 0&0&1&0&0&0
 \\0&0&0&0&0&1
 \\0&0&0&0&1&0
  \\0&1&0&0&0&0
 \\1&0&0&0&0&0
 \\
 \end {array}\right ), \ \ \
 {\bf A}_{(interaction)}^{(permuted)}=
 \left (
 \begin{array}{cc|cc|cc}
 Z&Y&G&B&F& B\\
 X&Z&C&F&C&G \\
 \hline
 F&B&Z&Y&G&B \\
 C&G&X&Z&C& F \\
 \hline
 G&B&F&B&Z &Y  \\
 C&F&C&G&X& Z
 \ea
 \right ).
 \label{sestidim}
 \ee
The partitioning of the latter asymmetric matrix indicates how our
system may be visualized as a coupled set of its three
two-dimensional asymmetric subsystems of the form~(\ref{trojka}).

For the second option with $L=2$ it is remarkable to notice that
while the former pseudo-parity remains asymmetric and, hence,
non-Hermitian, the latter matrix of coupling constants ${\bf A}$
appears to be, for some unknown reason, symmetric,
 \be
 {\bf r}_{(6,2)}^{(permuted)}=
 \left (\begin {array}{cccccc}
 0&0&1&0&0&0
 \\1&0&0&0&0&0
  \\0&1&0&0&0&0
 \\0&0&0&0&0&1
 \\
 0&0&0&1&0&0
 \\
 0&0&0&0&1&0
 \end {array}\right ), \ \ \
 {\bf A}_{(interaction)}^{(permuted)}=
 \left (
 \begin{array}{ccc|ccc}
 Z&X&X&C&D& G\\
 X&Z&X&G&C&D \\
 X&X&Z&D&G&C \\
 \hline
 C&G&D&A&B& B \\
 D&C&G&B&A &B  \\
 G&D&C&B&B& A
 \ea
 \right ).
 \label{dimsesti}
 \ee
Moreover, its inspection reveals that it is again tractable as a
coupled system of its two three-dimensional subsystems of the
symmetric-matrix form~(\ref{ctyrka}).

One has to add that the diagonalization of the $Z=0$  version of the matrix 
(\ref{sestidim}) is still feasible
and gives the two nondegenerate eigenvalues
 \ben
 F_{\pm 0} =
 G+R \pm \sqrt {2\,CY+2\,BX+XY+4\,BC}
 \een
and the two doubly degenerate eigenvalues
 \ben
 F_{\pm 1}=
 -\frac{1}{2}\,G-\frac{1}{2}\,R \pm \frac{1}{2}\,\sqrt
 {-3\,{G}^{2}+6\,GR-3\,{R}^{2}-4\,BX+4\,XY+4\,BC -4\,CY}.
 \een
In contrast, the study of the model
(\ref{dimsesti}) 
is hindered by the occurrence of the two different couplings on the main diagonal.
One must employ
 a shift of the eigenvalues
$F = \lambda + (Z+A)/2$ which leads to the diagonalization of the
matrix of the form
 \ben
\left (\begin {array}{cccccc}
-\omega&X&X&C&D&G\\\noalign{\medskip}X&-\omega&X&G
&C&D\\\noalign{\medskip}X&X&-\omega&D&G&C\\\noalign{\medskip}C&G&D&\omega&Y&Y
\\\noalign{\medskip}D&C&G&Y&\omega&Y\\\noalign{\medskip}G&D&C&Y&Y&\omega
\end {array}\right )
 \een
with non-vanishing main diagonal, $\omega=(Z-A)/2$. Still, in a
way resembling the previous $L=1$ model we get the two
nondegenerate eigenvalues
 \ben
 \lambda_{\pm 0} = Y+X \pm \sqrt{\Lambda_0}
 \een
with the abbreviation
 \ben
 \Lambda_0={Y}^{2}-2\,XY+{X}^{2}+{\omega}^{2}+
 \een
 \ben
 +2\,Y\omega-2\,X\omega+{D}^{2}
 +2\,DC+{C}^{2}+2\,GD+{G}^{2}+2\,GC,
 \een
accompanied by the two doubly degenerate eigenvalues
 \ben
 \lambda_{\pm 1}=
 -\frac{1}{2}\,Y-\frac{1}{2}\,X \pm \frac{1}{2}\,\sqrt{\Lambda_1}
 \een
where the discriminant ${\Lambda_1}$ is equal to the sum
 \ben
 {Y}^{2}-2\,XY+{X}^{2}+4\,{D}^{2}+4\,{\omega}^{2}-4\,Y\omega+
 \een
 \ben
 +4\,X\omega-4\,
 DC+4\,{C}^{2}-4\,GD-4\,GC+4\,{G}^{2}.
 \een

\section{Concluding remarks}

\subsection{Seven and more coupled channels at odd $K=2M-1$}

The pattern initiated by $K=3$ and $K=5$ is perpetuated at the
next odd dimension $K=7$ where the linear system of 49 equations
yields the four free parameters in the $L-$independent solution
 \be
 {\bf A}_{(interaction)}=
 \left (
 \begin{array}{ccccccc}
 Z&X&Y&D&D& Y&X\\
 X&Z&X&Y&D&D&  Y \\
 Y&X&Z&X&Y&D& D \\
 D&Y&X&Z&X& Y& D \\
 D&D&Y&X&Z &X &Y \\
 Y&D&D&Y&X& Z& X \\
 X &Y &D &D &Y& X& Z
 \ea
 \right ).
 \label{sedmidim}
 \ee
It is easy to see the general pattern and to guess the structure
of ${\bf A}_{(interaction)}$ at any higher odd dimension $K =
2M-1$. At any $M$ this hypothesis may be verified, say, in MAPLE,
by using the same algorithm as employed in the above calculations
performed at the first three nontrivial indices $M=2, \,M=3$ and
$M=4$.

In a way extrapolating the $K=5$ results we may move to $K=7$ in
eq. (\ref{sedmidim}) and guess the exceptional eigenvector $\left
\{1,1,1,1,1 ,1,1\right \}$ and its eigenvalue $F_0=
2\,D+2\,X+2\,Y$. The three further eigenvalues of the matrix
 \ben
 \left (\begin {array}{ccccccc}
 0&X&Y&D&D&Y&X\\
 X&0&X&Y&D&D&Y\\
 Y&X&0&X&Y&D&D\\
 D&Y&X&0&X&Y&D\\
 D&D&Y&X&0&X&Y\\
 Y&D&D&Y&X&0&X\\
 X&Y&D&D&Y&X&0
 \end {array}\right )
 \een
are doubly degenerate and may be written in the form of Cardano
formulae. They represent the real roots in terms of the complex quantities
and, for this reason, we omit them here and leave their generation to the interested 
readers as an easy exercise.

\subsection{Eight and more coupled channels at even $K=2M$}

One gets lost when solving eq. (\ref{hermiti}) in the Hermitian
case at $L=4$ with 37 free parameters in ${\bf A}_{8,4}$, 29 of
which occur there in pairs.

Similarly, not enough symmetry is induced by the $L=2$
pseudo-parity since the resulting asymmetric ${\bf A}$ depends on
as many as 16 free parameters, each occurring strictly four times.

Thus, the only satisfactory reduction is obtained from the
non-Hermitian pseudo-parity ${\bf r}_{(8,L)}$ with $L=1$ (and,
identically, $L=3$ etc). The explicit solution of the
corresponding eq. (\ref{hermiti}) leads to the permission of the
eight free coupling strengths, each occurring eight times in ${\bf
A}$. One parameter sits simply on the main diagonal and one on the
two ``submain" diagonals of the two off-diagonal quadrants. The
remaining six octets form the asymmetric pattern of three doublets
reflected pairwise by the main diagonal. The first pair jumps over
the main diagonal and the second one over the two submain
diagonals while the last pair simply fills the remaining
vacancies. We may again separate the even and odd rows and columns
and obtain a re-arranged matrix ${\bf A}$ which is partitioned in
the four submatrices where the two diagonal ones are composed of
the four parameters only and represent just a transposition of
each other. Each of the two off-diagonal blocks depends just on
the two parameters in a way resembling slightly the partitioned
structure of eq.~(\ref{dimctyr}).

\section*{Acknowledgement}

Work supported by the grant Nr. A 1048302 of GA AS CR.

%\end{document}

\newpage

\section*{Appendix A: A note on the interpretation of the Hamiltonians}

In a less usual approach to Quantum Mechanics as outlined in
review  \cite{Geyer} the Hamiltonian is allowed to be
quasi-Hermitian, i.e., Hermitian with respect to some
``anomalous", nontrivial metric $\Theta \neq I$ in the Hilbert
space of states. Of course, one must, first of all, re-define the
new, ``anomalous" scalar product
 \be
 |a\rangle \odot |b\rangle \,\equiv\,
 \langle a | \Theta | b \rangle\,, \ \ \ \ \
  \ \ \Theta = \Theta^\dagger>0.
  \label{onega}
 \ee
In effect, this is equivalent to a replacement of the standard
Hilbert space ${\cal H}$ by its new and $\Theta-$dependent
``physical" version ${\cal V}$ equipped with the more flexible and
adaptable product (\ref{onega}).

The key point is that once we stay within the innovated space
${\cal V}$, all the basic principles of Quantum Mechanics remain
unchanged. At the same time, the new flexibility carried by our
freedom of the choice of $\Theta$ is compensated by the loss of
the meaning of the standard Hermitian conjugation $H \to
H^\dagger$. Indeed, the standard textbook Hermiticity of the
observables $A=A^\dagger$ must be replaced by the requirement
 \ben
 |A\,a\rangle \odot |b\rangle \,\equiv\,
 |a\rangle \odot |A\,b\rangle
 \een
which may be re-read as an obligatory quasi-Hermiticity property
 \be
 A^\dagger = \Theta\,A\,\Theta^{-1}\,
 \label{quasi}
 \ee
of all the observables $A$ in the new formalism. Of course, no new
physics is being discovered in this manner because all the
operators $A$ with the property (\ref{quasi}) are simply Hermitian
with respect to the {\em fixed} new metric $\Theta\neq I$
\cite{Geyer} denoted usually as $\eta_+$ by Mostafazadeh
\cite{ali} and factorized as $\Theta={\bf CP}$ with ``charge"
${\bf C}$ by Bender et al~\cite{BBJ}.

\end{document}